\documentclass[twocolumn,aps,pre,showpacs,preprintnumbers,floatfix,amsmath,amssymb,superscriptaddress,longbibliography]{revtex4-1}

\usepackage{graphicx}    
\usepackage{bm}          
\usepackage[usenames,dvipsnames]{color}

\usepackage[colorlinks=true ,citecolor=blue]{hyperref}

\newcommand{\ee}{\mathrm{e}}
\newcommand{\ii}{\mathrm{i}}

\begin{document}	
\title{Quantum Transport Properties of an Exciton Insulator/Superconductor Hybrid Junction}

\author{D. Bercioux}
\affiliation{%
Donostia International Physics Center (DIPC), Manuel de Lardizbal 4, E-20018 San Sebasti\'an, Spain}
\affiliation{IKERBASQUE, Basque Foundation of Science, 48011 Bilbao, Basque Country, Spain}
\author{B. Bujnowski} 
\affiliation{Donostia International Physics Center (DIPC), Manuel de Lardizbal 4, E-20018 San Sebasti\'an, Spain}
\author{F. S. Bergeret}
\affiliation{Donostia International Physics Center (DIPC), Manuel de Lardizbal 4, E-20018 San Sebasti\'an, Spain}
\affiliation{Centro de F\'isica de Materiales (CFM-MPC) Centro Mixto CSIC-UPV/EHU,
20018 Donostia-San Sebastian, Basque Country, Spain}



\begin{abstract}
We present a theoretical study of electronic transport in a hybrid junction consisting of an excitonic insulator sandwiched between a normal and a superconducting electrode. The normal region is described as a two-band semimetal and  the superconducting lead as a two-band superconductor. In the excitonic insulator region, the coupling between carriers in the two bands leads to an excitonic condensate and a gap $\Gamma$ in the quasiparticle spectrum. We identify four different scattering processes at both interfaces. Two types of normal reflection, intra- and inter-band; and two different  Andreev reflections, one retro-reflective within the same band and one specular-reflective between the two bands. We calculate the differential conductance of the structure and show the existence of a minimum at voltages of the order of the  excitonic gap. Our findings are useful towards the  detection of  the excitonic condensate and provide a plausible  explanation of  recent transport experiments on  HgTe quantum wells and InAs/GaSb bilayer systems.
\end{abstract}

\maketitle
Particles condensates are at the basis of superconductivity and superfluidity~\cite{Leggett:2006,Leggett:2008}. We can explain superconductivity in terms of condensation of Cooper pairs ~\cite{Bardeen:1957}.  Similar processes of pairing and condensation can also take place  in other systems.  Mott postulated that semimetals (SMs) at sufficiently low temperatures could undergo a phase transition into an insulating state described by electron-hole bound pairs forming an exciton insulator (EI)~\cite{Mott:1961}.  Interestingly, the EI phase can be described by a BCS-like theory~\cite{Keldysh:1965,Jerome:1967}.
The coupling strength of an EI is expected to be even weaker than the coupling in a conventional superconductor (S) and very fragile with respect to disorder~\cite{Zittartz:1967}.  Additionally,  excitons may recombine quite fast, thus not allowing to form a condensate easily.
So far, the EI remains an elusive phase of matter since the original proposals~\cite{Jerome:1967,Zittartz:1967,Halperin:1968}. To date, one of the most successful attempts to obtain a stable EI is based on the condensation of excitons coupled to light confined within a cavity | the so-called exciton-polaritons~\cite{Kasprzak:2006}.

In order to reduce  the electron-hole recombination rate it was suggested to use a system where electrons and holes are spatially separated. Examples of this are  bilayer quantum well (QW) systems~\cite{Lozovik:1976,Shevchenko:1976},  topological EIs~\cite{Seradjeh:2009,Pikulin:2014,Budich:2014},  double two-dimensional electron gases in a strong magnetic field~\cite{Macdonald:1990,Yoshioka:1990,Eisenstein:2004, Spielman:2000,Kellogg:2004,Tutuc:2004},  two parallel, independently gated graphene monolayers separated by a finite insulating barrier~\cite{Lozovik:2011,Min:2008,Kharitonov:2008,Mak:2011} and very recently two bilayer graphene electron system separated by hexagonal BN or WSe$_2$~\cite{Li:2016,Lee:2016,Su:2017,Burg:2018}.

The fact that the flow of the EI condensate does not carry any charge current makes its detection rather difficult in bulk systems. Recent experiments suggested  signatures of the EI phase in TiSe$_2$ in the 1T phase~\cite{Monney:2010,Rossnagel:2011,Kogar:2017}. Also, metallic carbon nanotubes can be seen as SMs, thus are candidates to host an EI phase~\cite{Varsano:2017,Aspitarte:2017,Senger:2018}. There are also investigations of Ta$_2$NiSe$_5$ as a host material for the EI phase~\cite{Lu:2017,Mor:2017}.
However, results remain controversial. 

In this article, we propose a way for the detection of an EI condensate by electrical measurements.
Specifically,  we focus on a setup consisting of an SM  ``sandwiched" between a normal and a superconductor electrode [see Fig. \ref{figure1}(a)], which is similar to the one explored in Refs.~\cite{Kononov:2016,Kononov:2017}. These experiments showed the presence of a zero--bias anomaly in the differential resistance compatible with the size of the estimated EI gap of these systems. Both HgTe QWs and InAs/GaSb bilayer can be considered as SMs where the conduction band (CB) and the valence band (VB) have a finite energy overlap, but the relative minimum and maximum are displaced in the reciprocal space. 
Our results of the transport properties provide an interpretation of the experimental results and show that the zero-bias resistance can be associated with the existence of an EI condensate in the structure.

At a microscopic level the electronic transport through the structure shown in Fig.~\ref{figure1}(a)  result from the interplay between two channels of normal reflection and two channels of Andreev reflection.  One of the normal reflection channels is the standard intra-band specular reflection, whereas the second one is an intra-band normal reflection directly associated to the presence of the EI region with a length of the order of the associated EI coherence length. This additional reflection dominates in the case that the length of the EI region exceeds the coherence length of the condensate. 
The presence of the superconductor gives also  rise  to two different  Andreev reflections: a standard intra-band Andreev reflection that has a retro-reflective character as in standard  S/N systems, and an inter-band Andreev specular reflection, similar to the Andreev specular-reflection in  chiral SM~\cite{Beenakker:2006}. Contrary to the case of a chiral SM, the two channels of Andreev reflection can be open simultaneously, but the retro-reflective is usually dominating over the specular-one.

This Article is structured as follows: in Sec.~\ref{model} we present the model of a bulk EI coupled to a normal contact and to a superconductor contact; in Sec.~\ref{transport}, we analyse the quantum transport properties of this hybrid junction and identify the reflection processes mentioned above.  In Sec.~\ref{diff}, we present results for the differential conductance of a two terminal system, and we contrast them with the experiments on differential resistance in Ref.~\cite{Kononov:2016,Kononov:2017}.

\section{Model and Formalism}\label{model}
%
%
\begin{figure*}[!t]
\begin{center}
\includegraphics[width=.85\textwidth]{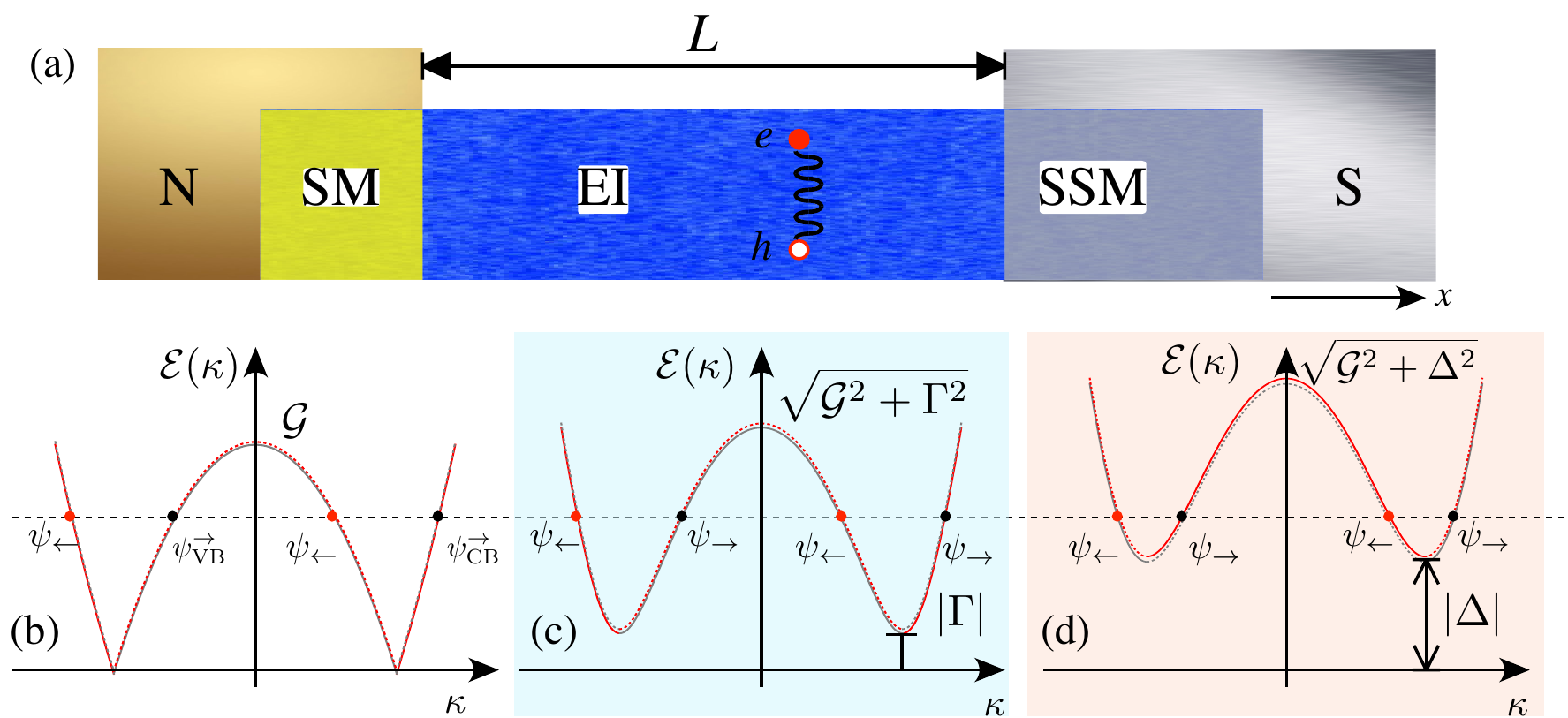}
\caption{\label{figure1} (a) A sketch of the normal/excitonic insulator/superconducting hybrid junction with a normal (N) electrode, the middle  EI region of length $L$, and a superconductive electrode on the right (S). (b)-(d) Energy spectra for the three regions, the semi metallic region [below the normal contact --- Panel (b)], the excitonic insulator region [between the contacts --- Panel (c)] and the superconducting one [below the superconducting contact --- Panel (d)]. In panel (b) we have: the band for an electrons in conduction band (solid red line), the band for a holes in the conduction band (dashed red line), the band for an electrons in the valence band (solid black line) and the band for a holes in the valence band (dashed black line). The same identification is used in panel (c) and (d), referring to the quasi-electron and quasi-hole  in the conduction and valence bands.}
\end{center}
\end{figure*}
%
%
We consider a bulk SM, where we neglect the displacement in the reciprocal space between the minimum of the CB and the maximum of the VB~\cite{Jerome:1967,Rontani:2005b}. The SM is contacted on the left with a normal metal and on the right with a superconductor. The two metallic contacts screen the Coulomb interaction in the regions close to them partially, thus, we can assume that Coulomb interaction is spatially modulated. In the unscreened region, we assume the formation of an EI characterised by a coupling strength of $\Gamma$.  Due to the proximity effect, a superconductive semi-metal (SSM) region is formed below the superconducting contact. An order parameter  $\Delta$ characterises the superconducting state. 
 
In other words, the structure shown in Fig.~\ref{figure1}(a) is modelled as a junction consisting of an SM electrode ($\Gamma=\Delta=0$),  a central region of length $L$ with a finite EI coupling ($\Gamma\neq0$ \& $\Delta=0$), and a SSM electrode with zero EI coupling and finite superconducting gap ($\Gamma=0$ \& $\Delta\neq0$). 
Overall, we account for a finite overlap $\mathcal{G}$ between the CB and VB. Formally, the hybrid junction is  described  by the  following matrix Hamiltonian:
%
%
\begin{equation}\label{Ham}
\mathcal{H}= 
\tau_z\otimes\left[\left( \frac{\bm{p}^2}{2m}-\mathcal{G}\right)\sigma_z + \Gamma(x)\sigma_x\right] + \Delta(x)\tau_x\otimes\sigma_0 \,.
\end{equation}
%
%
where $\bm{p}=-\ii \hbar \partial_{\bm{r}}$ is the standard momentum operator and the pairing functions are
%
%
\begin{subequations}\label{pairing}
\begin{eqnarray}
\Gamma(x)&=&\Gamma\Theta(L-x)\Theta(x)\label{Gamma}\,, \\
\Delta(x)&=& \Delta \Theta(x-L)\label{Delta}\,,
\end{eqnarray}
\end{subequations}
%
%
and  $\Theta(x)$ is the Heaviside step function.
The Hamiltonian \eqref{Ham} is written in terms of the Pauli matrices $\bm{\sigma}$ in the band sub-space: CB and VB, and the Pauli matrices $\bm{\tau}$  electron and hole (Nambu) subspace ~\cite{DeGennes:1999}. 
As a consequence the Hamiltonian is written in the basis defined by the \emph{bi}-spinor  $\Psi=~(\psi_{\mathrm{CB}},\psi_{\mathrm{VB}},\psi^\dag_{\mathrm{CB}},\psi^\dag_{\mathrm{VB}})$, ~\cite{Dolcini:2010,Peotta:2011,Bercioux:2017}.  The order parameters  $\Gamma$ and $\Delta$ are assumed to be step-like functions in real space and  no self-consistent calculation is carried out. 

We account for possible elastic reflections at the hybrid interfaces, SM/EI and EI/SSM,  by introducing two delta-like barriers in the  Hamiltonian Eq.~\eqref{Ham}~\cite{Blonder:1982,Rontani:2005b}: 
%
%
\begin{equation}\label{interfaces}
\mathcal{H}_\mathrm{int}=H_\mathrm{SM/EI}\delta(x)+H_\mathrm{EI/SSM}\delta(x-L)\,.
\end{equation}
%
%
These reflections can be ascribed, for example,  to the mismatch of the Fermi wave vector in the different regions.
Within this model, we assumed a sharp boundary between the different regions and an abrupt change of the excitonic and superconducting pair correlations at the SM/EI interface and EI/SSM one, respectively. This assumption is well justified if the size of the contact between the two regions is much smaller than its lateral dimensions.  This condition can be achieved experimentally by proper design of the contact. However, even in the case of a smooth transition of the pair correlations at the interface, we expect to have qualitatively similar results.

In the next three subsections, we introduce the scattering states necessary for evaluating the quantum transport properties of the hybrid junction. We assume incoming particles from the normal contact and determine the different scattering processes. 
\subsection{The semi-metal contact}

For the scattering state in the SM contact on the left side, we assume that both the CB and the VB will contribute to the overall conductance | this is justified by the equal population of both  bands at the Fermi energy~\cite{Jerome:1967} [c.f. Fig.~\ref{figure1}(b)]. Therefore, an incoming scattering state from the CB reads: 
%
%
\begin{subequations}\label{SM:states}

\begin{align}\label{SM:state:CB}
\psi_\mathrm{CB}=&\ee^{\ii q_y y} \left\{\frac{1}{\sqrt{k_+^\mathrm{SM}}}\left[\ee^{\ii  k^\mathrm{SM}_+ x}+ r_\mathrm{N-CB}\ee^{-\ii  k^\mathrm{SM}_+ x}\right]
\!\!\left(\begin{smallmatrix} 1 \\ 0 \\0 \\0 \end{smallmatrix}\right)  \right. \nonumber \\& +   \left.\frac{r_\mathrm{N-VB}}{\sqrt{k_-^\mathrm{SM}}}\ee^{\ii  k^\mathrm{SM}_- x}\left(\begin{smallmatrix} 0 \\ 1\\ 0 \\ 0 \end{smallmatrix}\right) 
+ \frac{r_\mathrm{A-CB}}{\sqrt{k_-^\mathrm{SM}}}\ee^{\ii k^\mathrm{SM}_- x} \left(\begin{smallmatrix} 0 \\ 0\\ 1 \\ 0 \end{smallmatrix}\right) \right. \nonumber \\&+ \left.\frac{r_\mathrm{A-VB}}{\sqrt{k_+^\mathrm{SM}}} \ee^{-\ii k^\mathrm{SM}_+ x} \left(\begin{smallmatrix} 0 \\ 0 \\ 0 \\ 1 \end{smallmatrix}\right)\right\}\,, 
\end{align}
%
%
whereas an incoming scattering state from the VB will read:
%
%
\begin{eqnarray}\label{SM:state:VB}
\psi_\mathrm{VB}&=&\ee^{\ii q_y y}\left\{ \frac{r'_\mathrm{N-CB}}{\sqrt{k_+^\mathrm{SM}}}\ee^{-\ii  k^\mathrm{SM}_+ x}
\left(\begin{smallmatrix} 1 \\ 0 \\0 \\0 \end{smallmatrix}\right)  \right. \nonumber \\ && \left. +\frac{1}{\sqrt{k_-^\mathrm{SM}}}\left[\ee^{-\ii  k^\mathrm{SM}_- x}+ r'_\mathrm{N-VB} \ee^{\ii  k^\mathrm{SM}_- x}\right]\left(\begin{smallmatrix} 0 \\ 1\\ 0 \\ 0 \end{smallmatrix}\right) \right. \nonumber \\&& 
\hspace{-1cm}+ \left. \frac{r'_\mathrm{A-CB}}{\sqrt{k_-^\mathrm{SM}}}\ee^{\ii k^\mathrm{SM}_- x} \left(\begin{smallmatrix} 0 \\ 0\\ 1 \\ 0 \end{smallmatrix}\right)+  \frac{r'_\mathrm{A-VB}}{\sqrt{k_+^\mathrm{SM}}} \ee^{-\ii k^\mathrm{SM}_+ x} \left(\begin{smallmatrix} 0 \\ 0 \\ 0 \\ 1 \end{smallmatrix}\right)\right\}\,.
\end{eqnarray}
\end{subequations}
%
%
In  Eqs.~\eqref{SM:states} we identify  four possible types of reflection: two normal reflections, where an electron is reflected as an electron either in the CB ($r_\mathrm{N-CB}$ and $r'_\mathrm{N-CB}$) or in the VB ($r_\mathrm{N-VB}$ and $r'_\mathrm{N-VB}$), and two  Andreev-like  reflections where an electron is converted in a hole either of the CB ($r_\mathrm{A-CB}$ and $r'_\mathrm{A-CB}$) or of the VB ($r_\mathrm{A-VB}$ and $r'_\mathrm{A-VB}$). 
In the two scattering states~\eqref{SM:states}, $k_\pm^\mathrm{SM}$ is the longitudinal component of the momentum.

\begin{figure*}[!t]
\begin{center}
\includegraphics[width=0.75\textwidth]{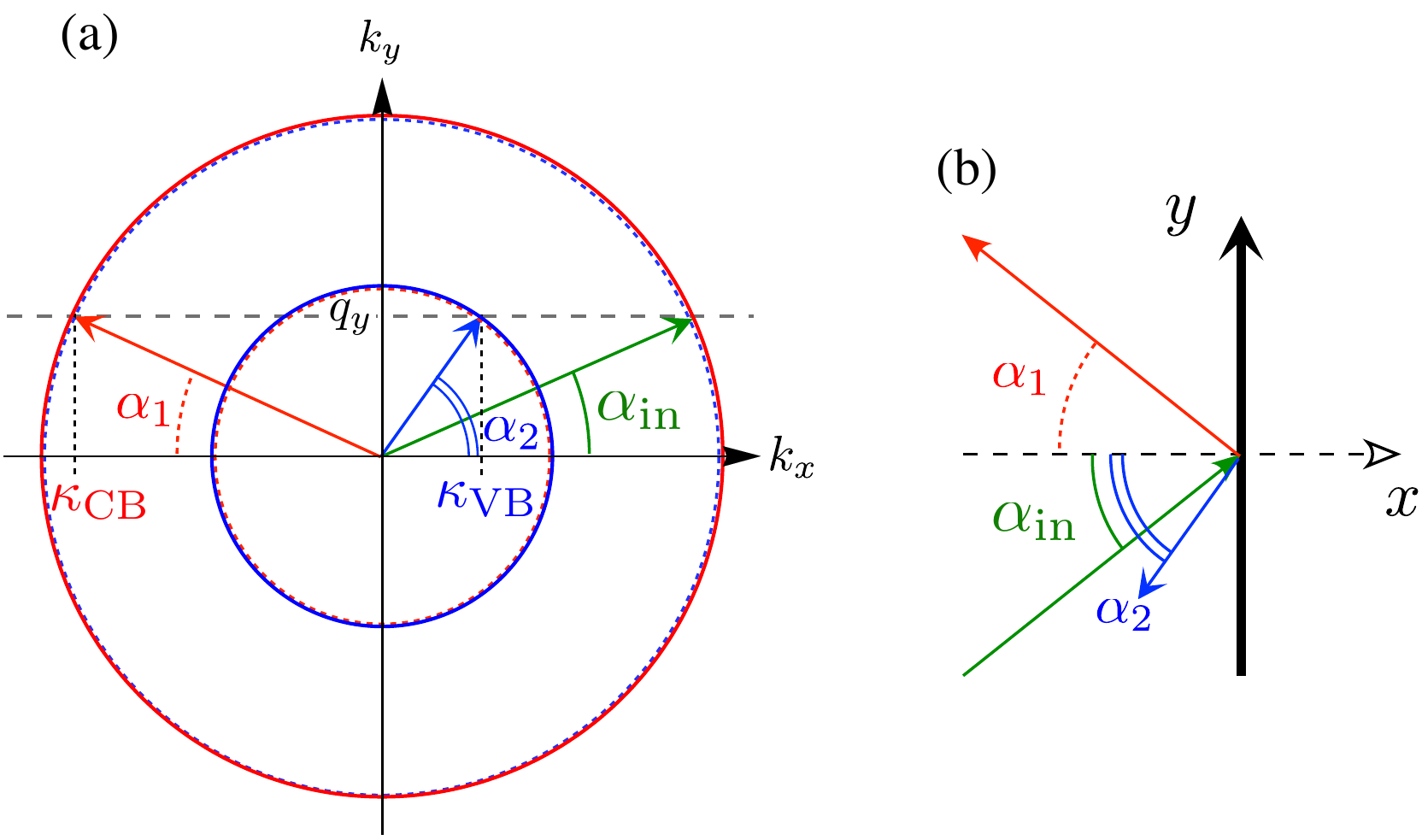}
\caption{\label{figure2} Panel (a): The two circles correspond  to the Fermi \emph{surface} for the CB  (in red) and the VB (in blue), respectively. The momentum along the $y$-axes $q_y$ is conserved in the scattering process. This produces reflection angles for normal reflection in the CB and for hole reflection in the VB $\alpha_1=\pi-\alpha_\mathrm{in}$. For the case of normal reflection in the VB and hole reflection in the CB, the reflection angle is  $\alpha_2$. Importantly, the directions of $\alpha_1$ and $\alpha_2$ are opposite. Panel (b): a sketch of the reflection angles with respect to the incoming one $\alpha_\mathrm{in}$.}
\end{center}
\end{figure*}

 The two Andreev reflections are qualitatively different: whereas  the Andreev reflection  within the same band is a retro-reflection process, 
 the inter-band Andreev process  is a specular reflection,  similarly to the one predicted in single- and bi-layer graphene~\cite{Beenakker:2006,Ludwig:2007,Efetov:2015}, Weyl SM~\cite{Chen:2013} and three-dimensional topological insulators~\cite{Breunig:2018,Bercioux:2018}.  The main difference is that in our hybrid junction, the two Andreev processes take place at the same energy, whereas in chiral SMs either it is either a retro- or a specular-reflection process. Also, the two types of  normal reflection have different propagation direction: 
 normal reflection within the same band is specular whereas it is retro-reflective when the band is changed (see Fig.~\ref{figure2} for more details).

The scattering dynamics can be understood considering the conservation of the energy  $\mathcal{E}$, and of the transversal momentum $q_y$. We can express the longitudinal component of the momentum $k_\pm^\mathrm{SM}$ in terms of these two conserved quantities:
%
%
\begin{equation}\label{momenta:SM}
k^\mathrm{SM}_\pm=\sqrt{\frac{2m}{\hbar^2}(\mathcal{G}\pm\mathcal{E})-q_y^2}\,.
\end{equation}
%
%
The injection of carriers from the VB~\eqref{SM:state:VB} is allowed only for energies smaller than the band overlap $\mathcal{G}$, than the VB closes and we are left with a single band system.

%
%
%
%
Using the same conservation considerations, we can obtain the various reflection angles (see Fig.~\ref{figure2}): 
%
%
\begin{subequations}\label{momenta:polar}
\begin{align}
k^{\mathrm{SM},\beta}_x&=\kappa_{\beta} \cos\alpha_\mathrm{in}\,,\\
q^{\mathrm{SM},\beta}_y&=\kappa_{\beta} \sin\alpha_\mathrm{in}\,,
\end{align}
%
%
with $\beta\in\{\mathrm{CB,VB}\}$. The moduli are expressed in terms of 
%
%
\begin{equation}
\kappa_\mathrm{CB/VB}= \frac{1}{\hbar}\sqrt{2m(\mathcal{G}\pm\mathcal{E})}.
\end{equation}
\end{subequations}
%
%
The intra-band normal reflection band has a propagation direction opposite to the incoming one; the same is true for inter-band Andreev reflection electrons that are converted into holes of the opposite band: $\alpha_1=\pi-\alpha_\mathrm{in}$.
For electrons injected from one band and converted into a hole of the same band (intra-band Andreev reflection), or into an electron of the other band (inter-band normal reflection), the reflection angle is:
%
%
\begin{equation}\label{alpha:special}
\alpha_2=\arcsin \left[ \frac{\hbar q_y}{\sqrt{2m(\mathcal{G}-\mathcal{E})}}  \right]\,.
\end{equation}
%
%

When the injection energy $\mathcal{E}$ exceeds the band overlap $\mathcal{G}$, the angle $\alpha_2$ becomes complex and the corresponding mode is evanescent. By imposing $2m/\hbar^2(\mathcal{G}-\mathcal{E})-q_y^2\le 0$ we can determine the critical injection angle $\alpha_\mathrm{c}$:
%
%
\begin{equation}\label{critical:alpha}
\alpha_\mathrm{c}=\arcsin\left[\sqrt{\frac{|\mathcal{G}-\mathcal{E}|}{\mathcal{G}+\mathcal{E}}}\right]\,.
\end{equation}
%
%

The intra-band Andreev reflection and the inter-band normal reflection are exactly retro-reflective only in the Andreev-limit of $\mathcal{G}\gg\mathrm{Max}[\Delta,\Gamma,\mathcal{E}]$, whereas by construction, the inter-band Andreev specular-reflection and the intra-band normal reflection are always opposite to the injection angle (a sketch of the various angles is in Fig.~\ref{figure2}). Furthermore, the inter-band Andreev reflection is a second-order process | it requires  scattering at both interfaces and a finite transmission  through the  EI region.

\subsection{The semi-metal superconducting contact}

In the SSM region, the superconducting pairing couples electrons and holes of the same band, but  we assume no inter-band  coupling.  An incoming particle from the SM region can be transmitted as quasi-electron or quasi-hole either in the CB ($t_\mathrm{qe-CB}$ and $t_\mathrm{qh-CB}$)  or in the VB  ($t_\mathrm{qe-VB}$ and $t_\mathrm{qh-VB}$). The transmitted wave function reads:
%
%
\begin{widetext}
{\small
\begin{align}\label{states:SSM}
\psi_\mathrm{SSM} = &\ee^{\ii q_yy} \left[t_\mathrm{qe-CB} \ee^{\ii k^\mathrm{SSM}_+ x} 
\left(\begin{smallmatrix}
u_\mathrm{CB} \\ 0 \\ v_\mathrm{CB} \ee^{\ii \phi}\\ 0
\end{smallmatrix}\right)
+ t_\mathrm{qh-CB} \ee^{-\ii k^\mathrm{SSM}_- x} 
\left(\begin{smallmatrix} 
v_\mathrm{CB} \\ 0 \\ u_\mathrm{CB}  \ee^{\ii \phi}\\ 0 
\end{smallmatrix}\right)
+t_\mathrm{qh-VB} \ee^{\ii k^\mathrm{SSM}_+x}
\left(\begin{smallmatrix}
0 \\ u_\mathrm{VB} \\ 0 \\ v_\mathrm{VB} \ee^{\ii \phi}
\end{smallmatrix}\right)
+t_\mathrm{qe-VB} \ee^{-\ii k^\mathrm{SSM}_- x}
\left(\begin{smallmatrix}
0 \\ v_\mathrm{VB} \\ 0 \\ u_\mathrm{VB} \ee^{\ii \phi}
\end{smallmatrix}\right)
\right]\,.
\end{align}}

%
%
The longitudinal component of the momenta in the SSM region [c.f. Fig.~\ref{figure1}(d)]  are
%
%
\begin{equation}\label{momenta:SSM}
k^\mathrm{SSM}_\pm=\sqrt{\frac{2m}{\hbar^2}(\mathcal{G}\pm\sqrt{\mathcal{E}^2-\Delta^2})-q_y^2}\,.
\end{equation}
\end{widetext}
%
%
The coherence factors, $u_\mathrm{CB/VB}$ and $v_\mathrm{CB/VB}$,  are the  ones of a $s$-wave superconductor.
Note that coherent factors have opposite signs for the CB and VB, this is due to the different curvature of these two bands:
%
%
\begin{subequations}\label{coherence:SSM}
\begin{align}
u_\mathrm{CB/VB} = &\sqrt{\frac{1}{2}\pm\sqrt{1-\left(\frac{\Delta}{\mathcal{E}}\right)}}\,,\\
v_\mathrm{CB/VB} = & \sqrt{\frac{1}{2}\mp\sqrt{1-\left(\frac{\Delta}{\mathcal{E}}\right)}}\,.
\end{align}
\end{subequations}
%
%
Importantly, we are assuming that the superconducting pairing is not pairing the CB and the VB; thus we are dealing with a two-band superconductor~\cite{Zehetmayer:2013}.

\subsection{The excitonic insulator}
In the middle region, there is a finite Coulomb pairing between electrons in the CB and the VB that leads to the excitonic insulator. 
 Here, we can express the wave function as counter-propagating states for the four possible states:
%
%
\begin{widetext}
\begin{align}\label{states:EI}
\psi_\mathrm{EI}(x) & =  \ee^{\ii q_yy}\left[\left(a_1 \ee^{\ii k^\mathrm{EI}_+ x} +b_1 \ee^{-\ii k^\mathrm{EI}_+ x} \right)
\left(\begin{smallmatrix}
u_e \\ v_e 
\\0 \\0
\end{smallmatrix}\right)+ \left(c_1 \ee^{-\ii k^\mathrm{EI}_- x} +d_1 \ee^{\ii k^\mathrm{EI}_- x} \right)
\left(\begin{smallmatrix}
v_e \\ u_e 
\\0 \\0
\end{smallmatrix}\right)\right. \nonumber \\
& \hspace{-0.7cm} \left.+\left(a_2 \ee^{\ii k^\mathrm{EI}_+ x} +b_2 \ee^{-\ii k^\mathrm{EI}_+ x} \right)
\left(\begin{smallmatrix}
\\0 \\0  \\ v_h \\ u_h 
\end{smallmatrix}\right)+ \left(c_2 \ee^{-\ii k^\mathrm{EI}_- x} +d_2 \ee^{\ii k^\mathrm{EI}_- x} \right)
\left(\begin{smallmatrix}
\\0 \\0  \\ u_h \\ v_h 
\end{smallmatrix}\right)\right]\,.
\end{align}
\end{widetext}
%
%

The longitudinal component of the momenta in the  EI region [c.f. Fig.~\ref{figure1}(c)] is defined as:
%
%
\begin{equation}\label{momenta:EI}
k^\mathrm{EI}_\pm=\sqrt{\frac{2m}{\hbar^2}(\mathcal{G}\pm\sqrt{\mathcal{E}^2-\Gamma^2})-q_y^2}\,,
\end{equation}
%
%
and the  coherence  functions are very similar to the superconducting one --- they read~\cite{Jerome:1967,Rontani:2005b,Rontani:2005}:
%
%
\begin{subequations}\label{coherence:EI}
\begin{align}
u_{e/h} & = \sqrt{\frac{1}{2}+\sqrt{1-\left(\frac{\Gamma}{\mathcal{E}}\right)}}\,,\\
v_{e/h} & = \sqrt{\frac{1}{2}-\sqrt{1-\left(\frac{\Gamma}{\mathcal{E}}\right)}}\,.
\end{align}
\end{subequations}
%
%

\section{ Transport properties of the hybrid junction}\label{transport}
We evaluate the  amplitudes of the various scattering processes by imposing the continuity of the wave functions 
at the two interfaces,  $x=\{0,L\}$: 
%
%
\begin{subequations}
\begin{align}
\psi_\mathrm{SM}(0^-) & = \psi_\mathrm{EI}(0^+)\,, \\
\psi_\mathrm{EI}(L^-) & =  \psi_\mathrm{SM}(L^+)\,.
\end{align}
\end{subequations}
%
%
Because of the delta-function  potential~\eqref{interfaces} at the interfaces, 
 the first derivative may exhibit a jump with a different sign  for the CB and VB electrons~\cite{Rontani:2005b,Rontani:2005}. Specifically, 
%
%
\begin{subequations}\label{BC}
\begin{align}
&\psi_\mathrm{EI}'(0^+)-\psi_\mathrm{SM}'(0^-)  = \frac{2mH_\mathrm{SM/EI}}{\hbar^2} \left(\tau_0\otimes\sigma_z\right)\psi_\mathrm{SM}(0) \label{BC:left}\,\\ 
&\psi_\mathrm{SSM}'(L^+)-\psi_\mathrm{EI}'(L^-)  = \frac{2mH_\mathrm{EI/SM}}{\hbar^2} \left(\tau_0\otimes\sigma_z \right)\psi_\mathrm{EI}(L)\label{BC:right}\,.
\end{align}
\end{subequations}
%
%

In order to simplify the discussion on the various scattering mechanisms, we consider the case of the injection of an electron from the CB, Eq.~\eqref{SM:state:CB}.
For energies smaller than $\Gamma$ and $\Delta$, at the first interface between the SM and the EI, the incoming electron can be normal reflected in the same band $r_\mathrm{N-CB}$ or normal reflected in the valence band $r_\mathrm{N-VB}$. The latter process is equivalent  to  the Andreev reflection at a normal metal-S interface~\cite{Rontani:2005b,Rontani:2005,Wang:2005,Bercioux:2017} and  leads  to the  formation of  an exciton pair.  

At the interface with the superconductor,  a quasi-particle state of the EI  condensate [a linear combination of CB and VB Eq.~\eqref{states:EI}] can be   Andreev-reflected  into a  hole either in the same band | $r_\mathrm{A-CB}$, or  into the  other band | $r_\mathrm{A-VB}$.

For energies higher than the EI gap, $\mathcal{E}>|\Gamma|$, the quasi-particles travel through the EI region as propagating waves. 
On the contrary, for energy below the excitonic gap,  the quasi-bound states are characterised by a complex momenta describing evanescent modes. 
The characteristic decay length of these  modes $\xi_\Gamma$ is energy dependent  and  is given by:
%
%
\begin{equation}\label{xiGamma}
\xi_\Gamma^{-1}=\sqrt{\frac{1}{2}\left(\chi_y+\sqrt{\chi_y^2+\frac{4m^2}{\hbar^4}(\Gamma^2-\mathcal{E}^2)}\right)}\,,
\end{equation}
%
%
where $\chi_y=q_y^2-\frac{2m\mathcal{G}}{\hbar^2}$; here $q_y$ is the momentum parallel to the interfaces that is conserved in the scattering process.  
When the band overlap $\mathcal{G}$ is the dominant energy scale, \emph{i.e.}, $\mathcal{G}\gg\max[\Gamma,\Delta,\mathcal{E}]$ | the Andreev limit~\cite{Andreev:1964,Zagoskin:2014} --- and therefore 
%
%
\begin{equation}\label{xiGammaApprox}
\xi_\Gamma=\frac{4\hbar\sqrt{m\mathcal{G}^3}}{\Gamma(4m\mathcal{G}+\hbar^2 q_y^2)}. 
\end{equation}
%
%

In the following we focus on the sub-gap transport,  \emph{i.e.} we consider the injection of  electrons from the SM contact with energies smaller than the superconducting gap, $\mathcal{E}<|\Delta|$; we further assume that $\Gamma<\Delta<\mathcal{G}$.
%
%
\begin{figure*}[!htb]
\begin{center}
\includegraphics[width=0.85\textwidth]{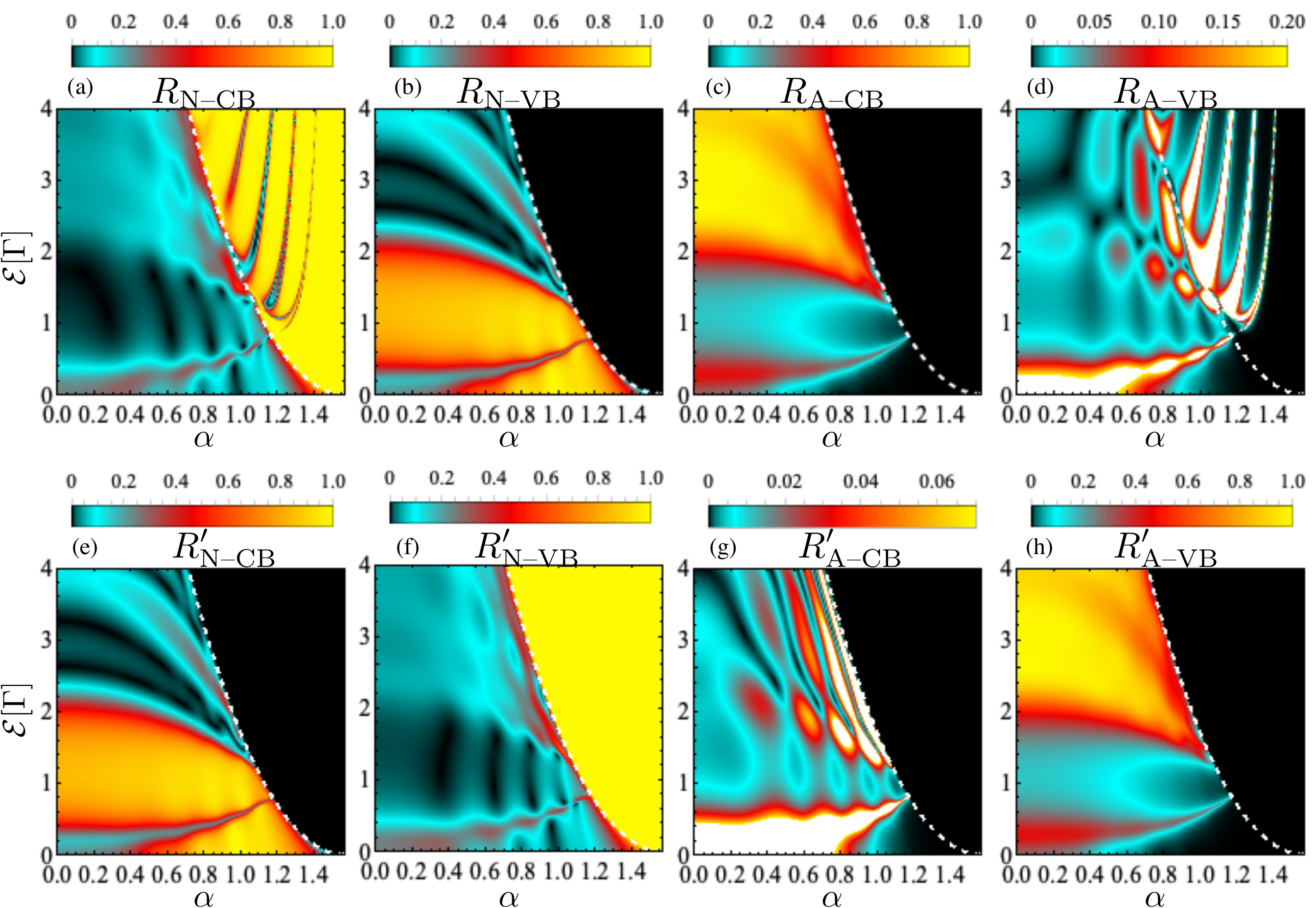}
\caption{\label{figure:three} The four reflection channels in the SM contact as a function of the injection energy and the injection angle. Panels (a) to (d): injection from the CB; Panels (e) to (h): injection from the VB. For the latter case, when $\alpha>\alpha_\mathrm{c}$ all the incoming electrons are reflected with probability 1 into the same contact, thus the remaining three reflection probabilities are equal to 0. For all the panels we  use $\mathcal{G}=10\Gamma$, $\Delta=4\Gamma$, $H_\mathrm{SM/EI}=H_\mathrm{EI/SSM}=0$ and $L=\xi_\Gamma$. In all the panels, the white-dashed line corresponds to the critical angle $\alpha_\text{c}$ defined in Eq.~\eqref{critical:alpha}.}
\end{center}
\end{figure*}
%
%

When injecting an electron from the CB, the probability for the four possible reflection channels in the SM electrode are then given by:
%
%
\begin{subequations}\label{probabilities:CB}
\begin{align}
R_\mathrm{N-CB}(\mathcal{E},\alpha) =&|r_\mathrm{N-CB}(\mathcal{E},\alpha)|^2\,, \label{NCB:CB} \\
R_\mathrm{N-VB}(\mathcal{E},\alpha) = &|r_\mathrm{N-VB}(\mathcal{E},\alpha)|^2 \Theta(\alpha_\mathrm{c}-\alpha) \label{NVB:CB}\,,\\
R_\mathrm{A-CB}(\mathcal{E},\alpha) = &|r_\mathrm{A-CB}(\mathcal{E},\alpha)|^2 \Theta(\alpha_\mathrm{c}-\alpha) \label{ACB:CB}\,,\\
R_\mathrm{A-VB}(\mathcal{E},\alpha) =&|r_\mathrm{A-VB}(\mathcal{E},\alpha)|^2\,, \label{AVB:CB}
\end{align}
\end{subequations}
%
%
where $\alpha$ is the injection angle. For injection angles larger than a critical  $\alpha_\mathrm{c}$, $R_\mathrm{N-VB}=R_\mathrm{A-CB}=0$ because of the lack of propagating electronic states in the VB or hole states in the CB, respectively. 

In contrast, when injecting from the VB, we need to account that for the angle of incidence $\alpha$ larger than the critical one, everything is reflected back into the VB and the other channels are closed:
%
%
\begin{subequations}\label{probabilities:VB}
\begin{align}
R'_\mathrm{N-CB}(\mathcal{E},\alpha) =&|r'_\mathrm{N-CB}(\mathcal{E},\alpha)|^2 \Theta(\alpha_\mathrm{c}-\alpha)\,,\label{NCB:VB} \\
R'_\mathrm{N-VB}(\mathcal{E},\alpha) = &|r'_\mathrm{N-VB}(\mathcal{E},\alpha)|^2\label{NVB:VB}\,,\\
R'_\mathrm{A-CB}(\mathcal{E},\alpha) = &|r'_\mathrm{A-CB}(\mathcal{E},\alpha)|^2 \Theta(\alpha_\mathrm{c}-\alpha) \label{ACB:VB}\,,\\
R'_\mathrm{A-VB}(\mathcal{E},\alpha) =&|r'_\mathrm{A-VB}(\mathcal{E},\alpha)|^2 \Theta(\alpha_\mathrm{c}-\alpha)\,.\label{AVB:VB}
\end{align}
\end{subequations}
%
%

The reflection amplitudes  in Eqs.~(\ref{probabilities:CB}-\ref{probabilities:VB}) are obtained by solving  the scattering problem numerically.

 Interestingly, by exchanging the injection band from CB to VB, the role on inter- and intra-band for normal and Andreev reflections get exchanged provided that the injection angle is smaller than the critical one $\alpha_\mathrm{c}$. This property is a consequence of the CB-VB  symmetry of the Hamiltonian~\eqref{Ham}. 
 
 It is also worth to notice the behaviour of the two Andreev channels when injecting from the conduction band. For injection angles larger than the critical one, the VB for electrons and the CB for holes are closed; thus, the only available channels for reflection are the normal one the specular Andreev one. The amount of normal and  specular Andreev reflections depend on the value of the band overlap $\mathcal{G}$: in fact, in the Andreev approximation, \emph{i.e.}, $\mathcal{G}\gg\mathrm{Max}[\Gamma,\Delta,\mathcal{E}]$, the critical angle~\eqref{critical:alpha} tends to $\pi/2$. In this limit, normal reflection and Andreev specular-reflection tend to zero.  In Fig.~\ref{figure:three} we summarise the properties we were describing above.

Moreover, in the Andreev approximation ($\mathcal{G}\gg\max[\Gamma,\Delta,\mathcal{E}]$) and  short junction limit $\delta=L/\xi_\Gamma\ll1$,  one can obtain analytical expressions for the reflection probabilities~\cite{Bercioux:2017}  at zero injection angle:
%
%
\begin{subequations}\label{seba:an}
\begin{align}
R_\mathrm{N-VB}=&4\delta^{2}\frac{\mathcal{E}^{2}\Gamma^{2}}{\Delta^{2}(\Gamma^{2}-\mathcal{E}^{2})}\,,\label{an:RRS} \\
R_\mathrm{A-CB}=&1-4\delta^{2}\frac{\mathcal{E}^{2}\Gamma^{2}}{\Delta^{2}(\Gamma^{2}-\mathcal{E}^{2})}\,,\label{an:RAR} \\
R_\mathrm{N-CB}=& 0\,,\\
R_\mathrm{A-VB}=& 0\,.
\end{align}
\end{subequations}
%
%
%
%
\begin{figure*}[!t]
\begin{center}
\includegraphics[width=0.85\textwidth]{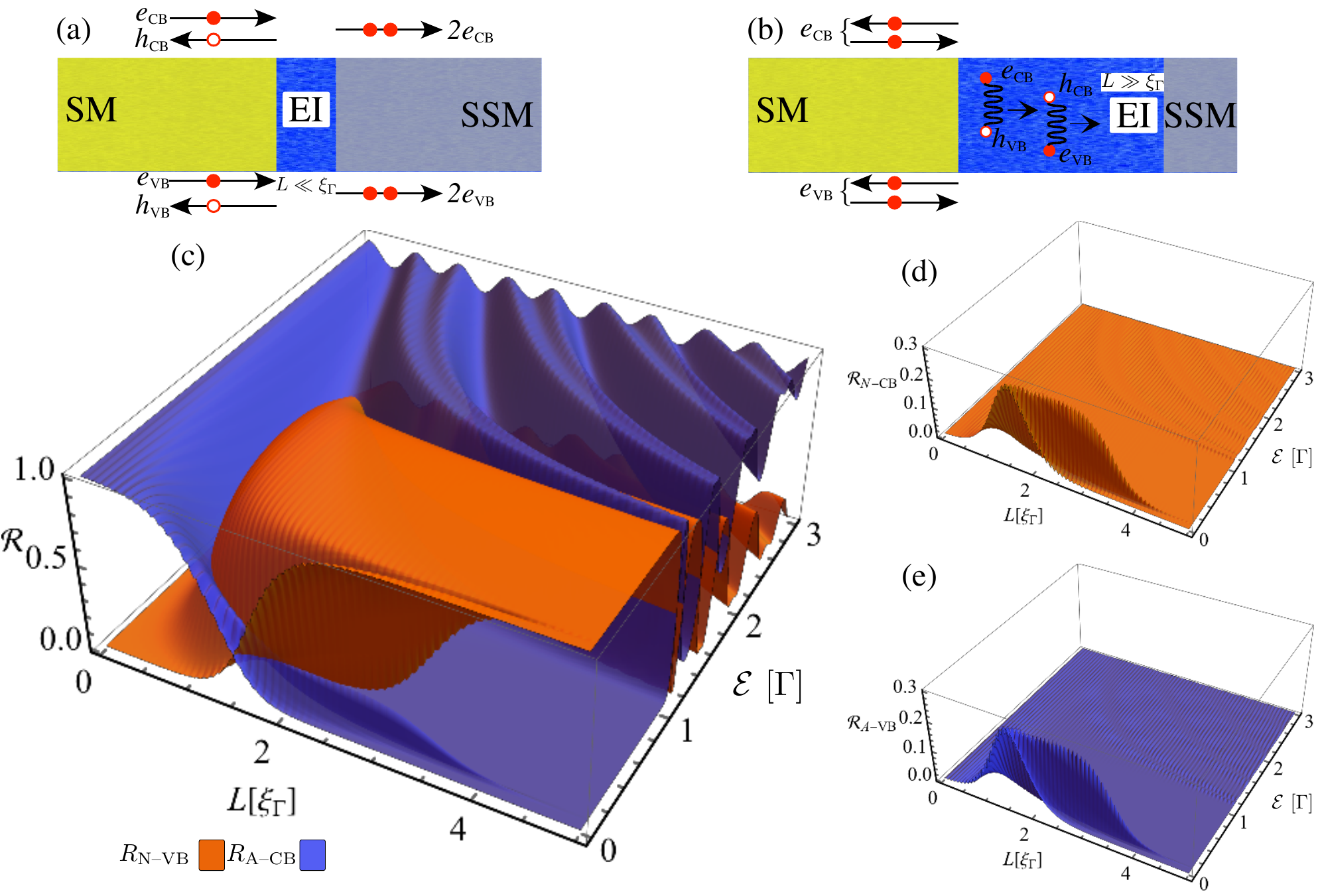}
\caption{\label{figure:four} Panel (a): sketch of the Andreev reflection with the CB or VB in the  short case, $L\ll \xi_\Gamma$. Panel (b): sketch of the normal reflection from CB to VB in the long junction limit  $L\gg \xi_\Gamma$. Panel (c): normal reflection $R_\mathrm{N-VB}$ and Andreev reflection $R_\mathrm{A-CB}$ as a function of the injection energy and length of the EI region. Panel (d): normal reflection $R_\mathrm{N-CB}$ as a function of the injection energy and length of the EI region. Panel (e): Andreev reflection $R_\mathrm{A-VB}$ as a function of the injection energy and length of the EI region. In panels (c) to (e) we use the following parameters: $\Delta=3\Gamma$ and $\mathcal{G}=20\Gamma$, $H_\mathrm{SM/EI}=H_\mathrm{EI/SSM}=0$.}
\end{center}
\end{figure*}
%
%

For an arbitrary length of the EI region,  we show the probability for the different reflection mechanism in  Fig.~\ref{figure:four}(c) to~\ref{figure:four}(e). In the limit  $L\rightarrow0$, the behaviour corresponds to a standard hybrid junction between an SM and a two-band superconductor. We expect that an incoming electron from the CB/VB is converted into a hole of the same band and no process with a change of band will take place~\cite{Blonder:1982,Bercioux:2017}. This behaviour is also exhibited if the length of the EI region is much shorter than the coherence length $\xi_\Gamma$:  in this case, the electron can tunnel through the EI region and reach the interface with the superconductor where it is converted into a respective hole.  This Andreev process is depicted in Fig.~\ref{figure:four}(a). 

In the other limiting, a large EI region $L\gg\xi_\Gamma$,  the junction behaves as a single interface between an SM and an EI~\cite{Rontani:2005b,Rontani:2005}. En electron injected from the CB in order to enter the EI region needs to form an exciton pair. Thus a hole from the opposite band is required~\cite{Rontani:2005b,Rontani:2005}. As a result, an electron from the other band is retracing the path of the incoming electron
[see Fig.~\ref{figure:four}(b)]. 

For arbitrary values of $L$  all  processes have a finite probability. Specifically, we have two second-order processes: the intra-band normal reflection and the inter-band Andreev reflection. Both require the presence of a finite length EI region. The latter is always weaker than the corresponding intra-band Andreev process.  Whereas inter-band normal reflection is always zero in the case of clean interface~\cite{Rontani:2005b,Rontani:2005}, it can be finite for a clean interface and a finite length $L$ of the EI region. In Fig.~\ref{figure:four}(c) we show the interplay between the inter-band normal reflection and the intra-band Andreev reflection as a function of the injection energy and the length of the EI region. In Figs.~\ref{figure:four}(d) and \ref{figure:four}(e) we show in the same fashion the intra-band normal reflection and the inter-band Andreev reflection. It is clear that for short and long EI regions, $R_\mathrm{N-VB}$ and $R_\mathrm{A-CB}$ are the dominating processes and that only at the crossover length when they exchange role the other two channels of reflection $R_\mathrm{N-CB}$ and $R_\mathrm{A-VB}$ get a finite value.

\section{Differential conductance and differential resistance}\label{diff}

In this section we use the previous results, Fig.~\ref{figure:three},  in order to describe the electronic transport properties of this hybrid junction. Our goal is to propose a direct electrical measurement scheme able to determine the size of the EI gap. 
At the end of the section, we provide a comparison with the experimental results by Kononov \emph{et al.}~\cite{Kononov:2016,Kononov:2017}.

\subsection{Electrical characterisation of the hybrid junction}

Given a finite transverse dimension $W$, the transverse momentum $q_y$ gets quantized accordingly to $q_y(n) = \frac{n\pi}{W}$
with $n\in \mathbb{N}$.
In general terms, the differential conductance~\cite{Blonder:1982} for an NS hybrid junction in an SM system can be defined as
%
%
\begin{align}\label{diffCond:0}
\frac{\partial I}{\partial V}(eV)  = &\frac{2 e^2}{h} \sum_{\beta\in\{\mathrm{CB/VB}\}}\sum_n \Big\{1-R_{\mathrm{N}-\beta}[eV,q_y(n)]\nonumber \\
&+R_{\mathrm{A}-\beta}[eV,q_y(n)]\Big\}\,,
\end{align}
%
%
where $R_{\mathrm{A}-\beta}[eV,q_y(n)]$ and $R_{\mathrm{N}-\beta}[eV,q_y(n)]$ are the Andreev and the normal reflection probabilities, respectively, and  with $\beta=\{\mathrm{CB/VB}\}$ indicating the band index. Equation~\eqref{diffCond:0} has a pre-factor $2e^2/h$ because of the two-fold spin degeneracy.
In the limit  of wide junctions, the spacing between different transverse modes can be considered negligible, and we can recast the sum into an integral over the (almost continuous) angle $\alpha_n\propto f[q_y(n)]$ using the following transformations:
%
%
\begin{align}\label{modetoangle}
\sum_n \to \frac{W}{2\pi}\int_{-\infty}^\infty d q_y & =  \frac{W}{2\pi}\int_{-\frac{\pi}{2}}^{\frac{\pi}{2}} \frac{\sqrt{2m(\mathcal{G+E})}}{\hbar}\cos\alpha d\alpha\nonumber \\ &=\frac{1}{2}\int_{-\frac{\pi}{2}}^{\frac{\pi}{2}} \mathcal{N}(\epsilon)\cos\alpha d\alpha\,,
\end{align}
%
%
where we have introduced $\mathcal{N}(\mathcal{E})=\frac{W}{\pi\hbar}\sqrt{2m (\mathcal{G+E})}$ as the number of open transversal modes at energy $\mathcal{E}$.

Within this approximation, we can express the differential conductance as:
%
%
\begin{align}\label{diffCond}
\frac{\partial I}{\partial V}  =& G_0(eV)\sum_{\alpha\in\{\mathrm{CB,VB}\}} \int_{0}^{\frac{\pi}{2}} \Big[1-R_{\mathrm{N-}\alpha}(eV,\alpha)\nonumber \\
&+R_{\mathrm{A-}\alpha}(eV,\alpha)\Big]\cos\alpha d\alpha\,,
\end{align}
%
%
with
%
%
\begin{equation}\label{G0}
G_0(\mathcal{E})=\frac{2e^2}{h}\mathcal{N}(\mathcal{E})
\end{equation}
%
%
being the differential conductance of the normal state. In writing  Eq.~\eqref{diffCond} we have used the  fact that all  reflection probabilities are an even function of the injection angle $\alpha$~\cite{Beenakker:2006,Bercioux:2018}. The conductance of the normal state will have an extra factor of $2$ due to the sum over the two open bands. 

In the Andreev approximation, we can use the expressions obtained for the short junction limit, Eqs. (\ref{seba:an}) 
in the case of a normal injection angle ($\alpha=0$), and  obtain 
%
%
\begin{equation}\label{an:DC}
\frac{\partial I}{\partial V}(eV) = 2G_0(eV)\left[ 1- 8 \delta^2 \frac{(eV)^2\Gamma^2}{\Delta^2(\Gamma^2-(eV)^2)} \right]\,.
\end{equation}
%
%
The factor $2$ is a result of the sum over the two open bands. From this expression, one sees the competition between a contribution stemming from the Andreev reflection that tends to increase the differential conductance and the appearance of the EI phase 
that suppresses it. 
 %
%
\begin{figure*}[!htb]
\begin{center}
\includegraphics[width=.85\textwidth]{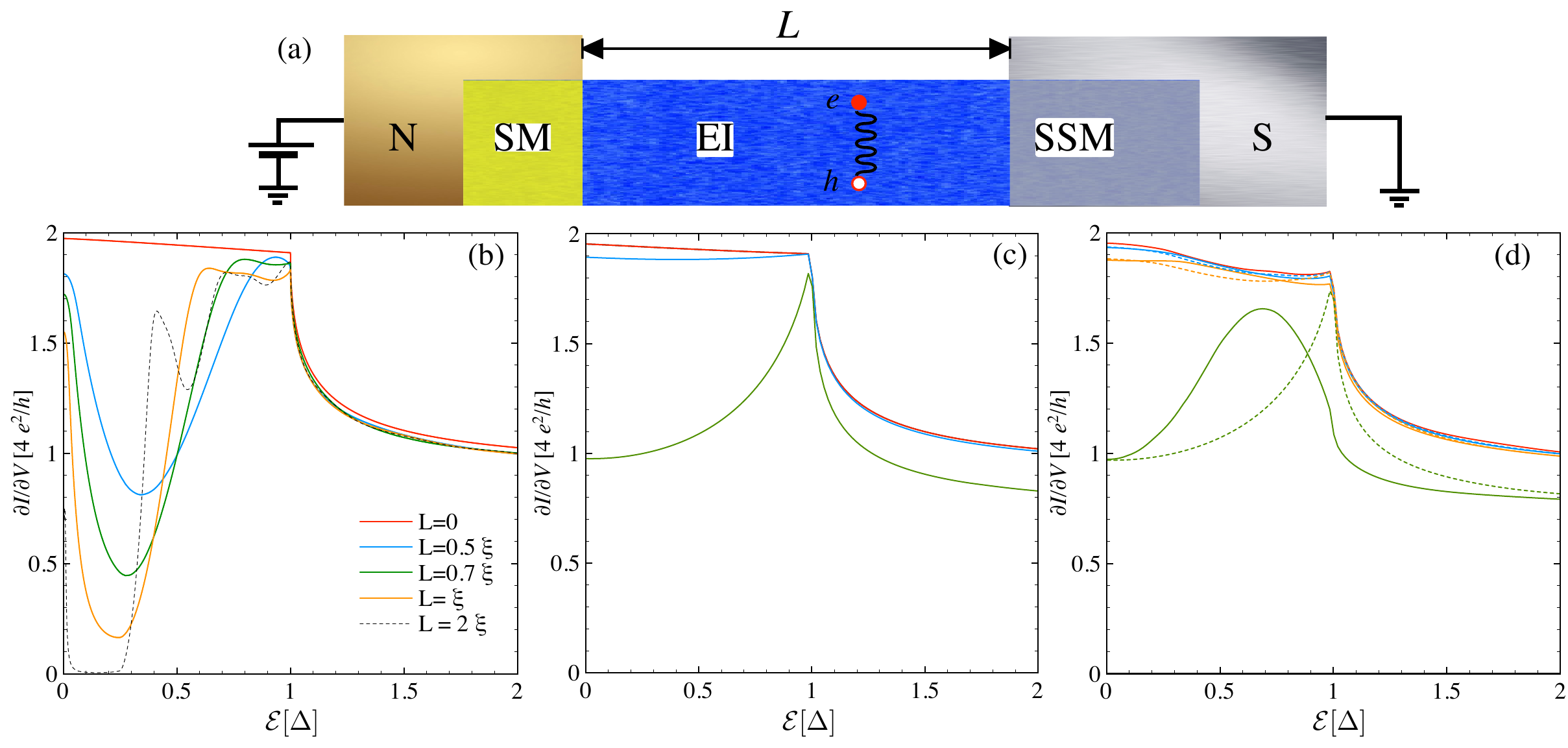}
\caption{\label{figure:five} Panel (a): Sketch of the electric configuration for measuring the differential conductance. (b) Differential conductance as a function of applied voltage for various length $L$ of the EI region, elastic reflection parameters equal to zero. (c) Differential conductance as a function of applied voltage for zero EI length and various strength of the elastic reflection potentials: $H=0.5$ (blue line), $H=1$ (orange line) and $H=5$ (green line). (d) Differential conductance as a function of applied voltage for finite EI length $L=0.1\xi_\Gamma$ and various strength of the elastic reflection potentials: $H_\mathrm{SM/EI}=0.5$ and $H_\mathrm{EI/SSM}=0$ (blue line), $H_\mathrm{SM/EI}=1$ and $H_\mathrm{EI/SSM}=0$ (orange line), $H_\mathrm{SM/EI}=5$ and $H_\mathrm{EI/SSM}=0$ (green line), $H_\mathrm{SM/EI}=0.$ and $H_\mathrm{EI/SSM}=0.5$ (blue dashed line), $H_\mathrm{SM/EI}=0$ and $H_\mathrm{EI/SSM}=1$ (orange dashed line), $H_\mathrm{SM/EI}=0$ and $H_\mathrm{EI/SSM}=5$ (green dashed line). In all panels $\mathcal{G}=100\Gamma$.}
\end{center}
\end{figure*}
%
%

We now focus on the numerical results for arbitrary junctions. 
The electric configuration for measuring the differential conductance is shown in Fig.~\ref{figure:five}(a), the system is configured in a way to ground the superconducting part and have a finite voltage only on the normal contact~\cite{Blonder:1982}.
We start by showing that the presence of a finite EI region via the additional inter-band channel of normal reflection strongly influences the behaviour of the differential conductance with respect to the injection energy. In Fig.~\ref{figure:five}(b), we show the differential conductance as a function of the injection energy for different lengths of the EI region. We observe that a first consequence of not having performed the Andreev approximation neither in the SM or the SSM contact is that the differential conductance is not constant~\cite{Blonder:1982} for energies smaller than the superconducting gap~\cite{Beenakker:2006,Bercioux:2018}. The most remarkable result is that for a finite length of the EI region the differential conductance shows a minimum. 
This minimum gets deeper and moves to lower energy by increasing the length of the EI region.
It is worth to note that this result is obtained in the absence of elastic reflections at the two interfaces $H_\mathrm{SM/EI}=H_\mathrm{EI/SSM}=0$. 

The standard behaviour of an SM/SSM interface without an EI region but with finite elastic reflection potentials $H_\alpha\neq0$ is shown in Fig.~\ref{figure:five}(c).  It is similar to the standard results obtained by Blonder \emph{el al.}~\cite{Blonder:1982}. Increasing the value of the $H$ parameter the overall differential conductance is decreased both for energies lower and higher than the gap $\Delta$.  This standard behaviour is very different from the one we have shown in Fig.~\ref{figure:five}(b) in the presence of the EI.

In Fig.~\ref{figure:five}(d) we show the differential conductance for $L=0.1\xi_\Gamma$.
We can observe that the effect of modulation of the differential conductance due to the presence of the EI is strongly enhanced only if the first interface potential ($H_\mathrm{SM/EI}>H_\mathrm{EI/SSM}$) is different from zero, whereas the second interface plays a minor role. 

\subsection{Comparison with the experiments}

In this section we use our model to interpret  experimental results of two recent works on  semimetal/superconductor hybrid junctions which 
suggest the existence of an  EI phase. 
 
We start discussing the experiment by Kononov \emph{et al.}~\cite{Kononov:2016}, where are investigated the  transport properties of a hybrid junction consisting of a HgTe QW.  The  width of the latter is  circa 20 nm and  from previous studies, the authors know that the system is an indirect SM with a small overlap between the bands~\cite{Kvon:2011}.  This property makes HgTe QWs reasonable candidates for observing the EI phase. The transport measurements showed a  zero bias anomaly in the differential resistance with an energy size comparable with the estimated EI gap of the system~\cite{Kononov:2016}. 
%
%
\begin{figure*}[!t]
\begin{center}
\includegraphics[width=0.85\textwidth]{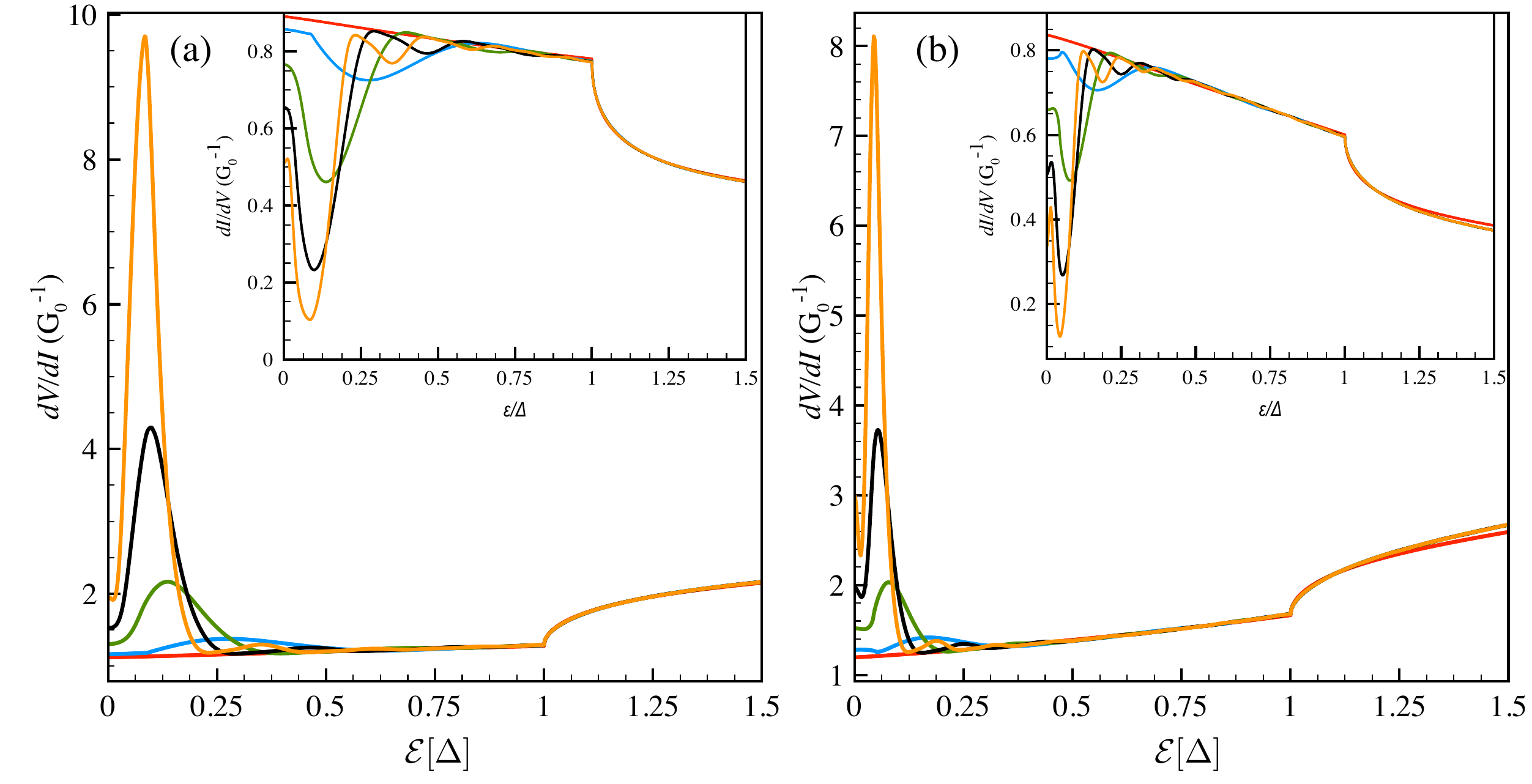}
\caption{\label{figure:six} Panel (a): Differential resistance of the hybrid structure  with a  Nb superconductive contact. In the inset we show the corresponding differential conductance. Panel (b): Differential resistance for the case of Nb/FeNi superconductive contact, in the inset we show the corresponding differential conductance. In both panels the various lines correspond to different length of the EI region: $L=0\xi_\Gamma$ (red line), $L=0.25\xi_\Gamma$ (blue line), $L=0.5\xi_\Gamma$ (green line), $L=0.75\xi_\Gamma$ (black line), and $L=\xi_\Gamma$ (orange line). The other parameters used in the plots are given  in the main text.}
\end{center}
\end{figure*}
%
%
As we show next the zero bias anomaly observed experimentally is compatible with the results of the last section.  In order to use our model, we need to assume that the momentum displacement between the CB and the VB is negligibly small. To evaluate the differential resistance, we take the inverse of the differential conductance in Eq.~\eqref{diffCond}. 
We use the values  of the system parameters given in Refs.~\cite{Kvon:2011,Kononov:2016} and set $\mathcal{G}=3.0$~meV, $\Gamma=0.05$~meV,  $\Delta_\mathrm{Nb}=1.15$~meV for the Nb/HgTe structure and $\Delta_\mathrm{Nb/FeNi}=0.6$~meV which should correspond to a (Nb/FeNi)/HgTe structure explored  in  Ref.~\cite{Kononov:2016}. We also assume a  finite transparency of the interface barriers $H_\mathrm{SM/EI}=H_\mathrm{EI/SSM}=0.5$~\cite{Kononov:2016}.
 For these parameters value, we have an EI coherence length of $\xi_\Gamma=955$~nm.

The results of the differential resistance are shown in the two panels of Fig.~\ref{figure:six} for the two different superconductor used in Ref.~\cite{Kononov:2016}. The solid red line describes an SM/SSM junction ($L=0 \xi_\Gamma$), and coincides with the well-known results of the Blonder-Tinkham-Klapwijk theory after integration over injection angle~\cite{Blonder:1982,Mortensen:1999,Beenakker:2006}. It is important to note that the ratio $\mathcal{G}/\Delta$ is of 2.6 and 5 for Nb (Panel a) and Nb/FeNi (Panel b), respectively. 
Thus the differential resistance at zero voltage is not exactly equal to $0.5 G_0^{-1}$, which would be the result obtained in the leading order when $\mathcal{G}/\Delta\gg1$.  
By increasing the length of the EI region, we obtain a peak in the differential resistance at energies smaller than $\Delta$.  These results resemble the experimental observations of  Ref.~\cite{Kononov:2016} and can  be attributed as a manifestation of the EI phase.

\begin{figure*}[!t]
\begin{center}
\includegraphics[width=.85\textwidth]{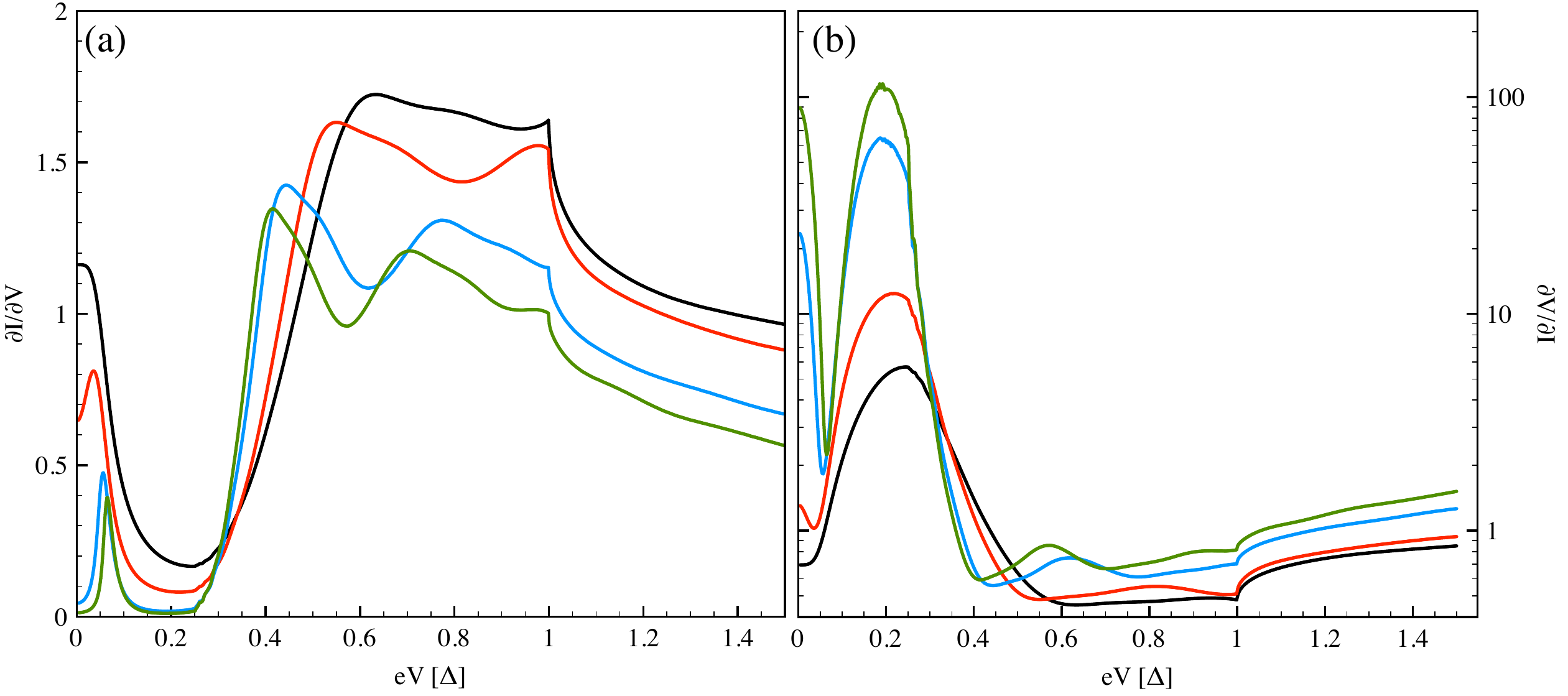}
\caption{\label{figure:seven} Panel (a): differential conductance as a function of the applied voltage; Panel (b) differential resistance as a function of the applied voltage. In both panels, the ratio between band overlap and superconducting gap is fixed at the following values $\mathcal{G}/\Delta=7.5$ (blue line), $\mathcal{G}/\Delta=5$ (red line), $\mathcal{G}/\Delta=2.5$ (black line), and $\mathcal{G}/\Delta=2$ (green line). For all the curves the length of the EI region is fixed equal to the coherent length for the black curve, $L=\xi_\Gamma[\mathcal{G}=30\Gamma]$, c.f. Eq.~\eqref{xiGamma}, and the superconducting gap is fixed to $\Delta=4\Gamma$.}
\end{center}
\end{figure*}

 Another system that may  host an EI phase are  InAs/GaSb double QWs. Here, electrons are located in the InAs QW and holes are located in the GaSb QW. Recent theoretical calculations~\cite{Xue:2018} confirmed the experimental observation that the EI phase exists in this system~\cite{Du:2017,Yu:2018}. Interestingly,  the bands  overlap,  can be changed  by changing the width of one of the QWs --- this change corresponds to a modulation of  $\mathcal{G}$ in our model. We have investigated the response of the system to changes of the band overlap in Fig.~\ref{figure:seven}. 
 
 Interestingly, we find at low-voltages two type of behaviors for  the conductance: the one discussed previously showing a minimum in the conductance [compare black line in Fig.~\ref{figure:seven} with  Fig.~\ref{figure:six}(a)], and  one with a clear maximum at finite voltage  (all the curves except than black in Fig.~\ref{figure:seven}). 
These two types of behavior resemble these observed recently  by Kononov \emph{at al.}~\cite{Kononov:2017}, who measured the differential resistance of  a superconducting  InAs/GaSb double QW junction for different with of the InAs gas.  In their explanation,  the different behaviors of the  differential resistance is associated with  the variation of the CB/VB overlap  and with presence of proximitized superconducting region with a smaller gap $\Delta'<\Delta$.  
 When changing from large $\mathcal{G}$ to a small one, the differential conductance shows a peak at certain finite voltage. Moreover, as in the experiment of Ref.~\cite{Kononov:2017}, the measured differential resistance  of the two  observed behaviors [black and green curves in Fig.~\ref{figure:seven}(b)] differ by a factor larger than 20. This could also be explained from  our assumption that an EI is formed between the electrodes. 

%
%
%
%
 
It is important to stress that in our model we neglect any kind of disorder.  One can expect  that disorder will suppress all the effects described above for the simple reason that  elastic disorder has  for the EI condensate~\cite{Zittartz:1967} the same effect as a low concentration of magnetic impurities has for conventional superconductivity, leading to a suppression of it.  Similarly, an enhancement of  temperature will wash out the non-monotonic behavior predicted for the differential conductance, as it has been seen experimentally~\cite{Kononov:2017}.
The main experimental features have been observed at the base temperature of 30 mK \cite{Kononov:2016,Kononov:2017}, thus justifying our approach of zero temperature.

In short, our results provide a first plausible qualitative explanation for the low-bias transport properties observed in the experiments in  Refs.~\cite{Kononov:2016,Kononov:2017}.
It is however  worth to mention that in addition to the low  bias anomalies described above, the experiments of  Ref.~\cite{Kononov:2016,Kononov:2017}  shown also features in the  sub-gap conductance that resembles the multiple-Andreev reflections  processes in voltage biased Josephson junctions~\cite{Averin:1995}. In principle such sub-gap features are unexpected, since in these experiments  there is only one superconducting electrode. Further research is needed to understand such sub-gap behavior.

\section{Conclusions \& Discussion}

We analyse the transport properties of a hybrid junction created in a semi-metallic system. We assume that one side of the system can be considered as a normal electrode, and the other side gets proximitized by a standard $s$-wave superconductor. In the region between these two contacts, we consider a finite  Coulomb interaction between electrons of the two bands that gives rise to an excitonic insulator. We show that the presence of the excitonic insulator can modify the transport properties of the system drastically. 

Specifically, we analyse the presence of four types of reflection in the normal electrode, two normal and two Andreev reflections. 
We have two intra-band processes: a normal specular reflection channel and an Andreev retro-reflection one. Besides, we have also two inter-band processes: a normal retro-reflection and an Andreev specular-reflection.

The interplay of these four channels of reflections has a strong influence on the electrical properties of the system.  If the length of the excitonic region is comparable to or larger than the corresponding coherence length, we observe a minimum in the differential conductance. This minimum is a  signature of the presence of an EI condensate in the system. We  propose  our  set-up 
for the detection of an EI condensate and we use our results for interpreting recent transport experiments in semi-metallic systems sandwiched between a normal and a superconducting contact~\cite{Kononov:2016,Kononov:2017}.

The results we obtained are similar to the results we obtained in Ref.~\cite{Bercioux:2017}, however, the physical contest is different, here we have considered a bulk semi\-metal system, whereas in Ref.~\cite{Bercioux:2017} we considered a bilayer system composed of two parallel two-dimensional electron gases that are hosting an excitonic insulator phase. The circuit we presented in Ref.~\cite{Bercioux:2017} for the detection of the excitonic insulating phase suffers from the problem of not being consistent with the survival of the excitonic insulating phase~\cite{Su:2008}.

\begin{acknowledgments}
Discussions with J. Cayssol, V. Golovach,  T.M. Klapwijk and M. Rontani  are acknowledged. 
The work of DB, BB, and FSB is supported by Spanish Ministerio de Econom\'ia y Competitividad (MINECO) under the project  FIS2014-55987-P, and by the Spanish Ministerio de Ciencia, Innovation y Universidades (MICINN) under the project FIS2017-82804-P, and by the Transnational Common Laboratory \emph{QuantumChemPhys}. 
\end{acknowledgments}

\bibliography{biblio}

\end{document}